\documentclass[aps,amsfonts,nofootinbib,twocolumn]{revtex4}
\usepackage{graphicx}
\usepackage{amsmath}
\def\be{\begin{equation}}
\def\ee{\end{equation}}
\def\bea{\begin{eqnarray}}
\def\eea{\end{eqnarray}}

\begin{document}

\title{Regular collision of dilatonic inflating branes}
\author{Emily Leeper, Kazuya Koyama and Roy Maartens
\vspace*{0.2cm}  }
\affiliation{Institute of Cosmology \&
Gravitation, University of Portsmouth, Portsmouth, PO1~2EG, UK
\vspace*{0.2cm}}
\date{15/08/05}
\begin{abstract}We demonstrate that a two brane system
with a bulk scalar field driving power-law inflation on the branes
has an instability in the radion. We solve for the resulting
trajectory of the brane, and find that the instability can lead to
collision. Brane quantities such as the scale factor are shown to
be regular at this collision.  In addition we describe the system
using a low energy expansion.  The low energy expansion accurately
reproduces the known exact solution, but also identifies an
alternative solution for the bulk metric and brane trajectory.
\end{abstract} \maketitle

\section{Introduction}

In the search for a theory that can describe both quantum
mechanics and gravity, string theory (which describes matter as
vibrating strings moving in 11 dimensions) seems a strong
candidate.  It has been the subject of much research, and many
different realisations of string theory have been found to be
possible.  This being so, experimental constraints on the theory
are vital.  It is hoped that the next generation of particle
accelerators now being built will reach energies where some sign
of extra dimensions may be seen, while laboratory gravity
experiments at low energies can put constraints on the length
scale of any extra dimensions.  In cosmology, we can probe both
the highest energies, through understanding of the early universe,
and the largest scales available to us; thus an understanding of
how string theory and extra dimensions affects cosmology is very
valuable.

The fact that we observe only three spatial dimensions has to be
explained away in string theory - this can be done by
compactifying the extra dimensions to very small scales, or by
attaching all the strings corresponding to normal matter to a
four-dimensional surface called a brane.  Five-dimensional models
based on this latter method of dimensional reduction were
presented in two papers by Randall and Sundrum in 1999 \cite{RS}.
Much work has been done on cosmological generalisations of these
models since then. Early universe inflation in braneworlds has
been extensively studied, mostly driven by some kind of inflaton
field, either localised on the brane with the other matter
fields~\cite{Maartens:1999hf}, or living in the whole
bulk~\cite{Himemoto:2000nd,Kobayashi:2000yh}. In this paper we
will be considering a model of bulk-driven inflation.

In \cite{koyama}, Koyama and Takahashi found an exact solution for
a five-dimensional bulk metric where a bulk scalar field with
non-BPS exponential potential drives power-law inflation on a
single brane. This scenario is particularly attractive as the
cosmological perturbations, normally an intractable problem in
braneworld scenarios, can be solved exactly \cite{koyama2}.

In \cite{Mukohyama&Coley}, Mukohyama and Coley developed the
scenario to include two branes.  They found that for any value of
the brane tension on the second brane there is one location where
the second brane will remain at a constant bulk coordinate away
from the first brane. This is somewhat analogous to the scenario
with two de Sitter (dS) branes and an empty Anti-de Sitter (AdS)
bulk studied by Gen and Sasaki \cite{dS2}. However unlike the
static de Sitter model, the presence of the scalar field means
that the constant coordinate separation does not lead to a static
equilibrium as the proper distance between the two branes is time
dependent.

The stability of this two-brane model against a wide class of
metric perturbations was shown in \cite{Mukohyama&Coley}. Although
it was shown that there were no unstable bulk metric
perturbations, including perturbations of the bulk metric
introduced by radion fluctuations, the radion itself was not
calculated.  Thus a scenario with a brane moving in the fixed bulk
space-time was not excluded by the analysis of
\cite{Mukohyama&Coley}.  Such a scenario would be the
natural analogue of the radion instability found in the two-dS
brane model in \cite{dS2}. In Section \ref{sec:zdot} we will
examine the case where the second brane is allowed to move in the
fixed bulk metric, and solve the junction conditions giving its
trajectory. We find that the second brane position is indeed
unstable, with any displacement leading to the second brane going
to infinity or colliding with the reference brane.  However, it is
shown that all brane quantities are completely regular at this
collision. This is in contrast with the results of Webster and
Davis \cite{websterdavis}, who examined similar non-BPS braneworld
models with more general scalar field potentials, and found that
the scalar field typically becomes divergent at a collision.

The stability of this scenario to homogenous metric perturbations
is examined in a low energy expansion in Section \ref{sec:lee}.
These results reinforce the stability found in
\cite{Mukohyama&Coley}. We also demonstrate the ability of the low
energy expansion to reproduce a known exact solution in the
correct limit.

\section{Exact solution}

\subsection{Previous Results}\label{sec:prevres}

The scalar field $\Phi$ in the bulk has potential \cite{koyama}:
\be\Lambda(\Phi)=\kappa^{-2}\left(\frac{\Delta}{8}+\delta
\right)\sigma_0^2e^{-2\sqrt{2}b\kappa\Phi(z,t)} \, ,
\label{bulkpotential}\ee and induced potentials on the two branes:
\be\lambda_{0,1}(\Phi)=\sqrt{2} \kappa^{-2} \sigma_{0,1}e^{-\sqrt{2}b\kappa\Phi(z,t)}\:
, \label{scalarfieldpotentialbr}\ee where $\Phi$ is taken to have
a separable form: \be \Phi(z,t)=\phi(t)+\Xi(z) \; . \ee The
parameter $\Delta$ is defined in terms of the coupling constant
$b$ as \be \Delta=4b^2 - \frac{8}{3} \; . \ee Then the part of the
bulk potential proportional to $\Delta$ is the BPS potential, and
tuned to the brane potentials such that the metric induced on the
branes will be static (this is equivalent to the Randall-Sundrum
tuning in the model with AdS bulk).  The parameter $\delta$ is a
free parameter giving the deviation from this BPS tuning, and is
what allows the time-dependent evolution on the brane.

The action for this model is given by

\bea S=&&\int d^5x
\sqrt{-g}\left[\frac{1}{2 \kappa^2}R-\frac{1}{2}\partial_{\mu}\Phi\partial^{\mu}\Phi-\Lambda(\Phi)
\right]
\label{action} \\
&&- \int d^4x \sqrt{-q_0}\lambda_0(\Phi) - \int d^4x
\sqrt{-q_1}\lambda_1(\Phi) \notag \, .  \eea Taking the bulk
metric ansatz: \be ds^2
=e^{2W(z)}\left[-dt^2+e^{2\alpha(t)}d\vec{x}^2 +
e^{2\sqrt{2}b\kappa\phi(t)}dz^2\right]\; ,\ee the Einstein
equations are solved by \cite{koyama}: \bea
e^{\alpha(t)}&=&(H_0t)^{2/(3\Delta+8)} \label{exactmetric1}
\\*[2mm] e^{\sqrt{2}b\kappa\phi(t)}&=&H_0 t \label{exactmetric2} \\*[2mm]
e^{W(z)}&=&\mathcal{H}(z)^{2/3(\Delta+2)} \label{exactmetric3}
\\*[2mm] e^{\kappa \Xi(z)}&=&\mathcal{H}(z)^{2\sqrt{2}b/(\Delta+2)} \label{exactmetric4} \eea
with \be H_0 = -\frac{3\Delta+8}{3(\Delta+2)}H \, , \quad
H=-(\Delta+2)\sqrt{-\frac{\delta}{\Delta}}\sigma_0  \ee and for
$\Lambda<0$, $\mathcal{H}(z)$ is given by: \be \mathcal{H}(z) =
\sqrt{-1-\frac{\Delta}{8\delta}}\sinh{Hz} \label{mathcalH}\ee

We assume that $-8/3<\Delta<-2$ so that all exponents are finite,
and that $\Delta/8 + \delta<0$ to ensure $\Lambda<0$. Then we note
that for $\delta\neq0$ a brane at position $z=z_0$, determined by
$\mathcal{H}(z_0)=1$, undergoes power law inflation with scale
factor $a(t)=(H_0t)^{2/(3\Delta+8)}$; and for a brane at this
location bulk time $t$ coincides with brane proper time. In the
limit $\delta\rightarrow0$ we have that $H=H_0=0$, giving a scale
factor of $0$, demonstrating that the power law inflation is
driven by the scalar field and by the $\delta$ part of the scalar
field in particular.

A second brane may be introduced at arbitrary constant $z=z_1$ as
in  \cite{Mukohyama&Coley}.  Proper distance between the branes is
then given by \be L=H_0t\int_{z_0}^{z_1}e^{W(z)}dz \, \propto t \;
. \ee  In the case of two dS branes the equilibrium position has
constant proper distance; the time-dependence of the power-law
inflation on the brane causes a simple time-dependence for the
equilibrium proper distance.  As $H_0$ is positive, once the
branes have emerged from the curvature singularity at $t=0$ they
are always moving apart.

The junction conditions at the first brane are automatically
satisfied; at the second brane they give a single equation
relating $z_1$ and the two brane tensions, $\sigma_0$ and
$\sigma_1$: \be \sigma_1 =
-2\sqrt{2}\sigma_0\sqrt{-1-\frac{\Delta}{8\delta}}\sqrt{\frac{-\delta}{\Delta}}\cosh{Hz_1}
\label{tuningreln} \, . \ee In proper time on the second brane the
scale factor gives power law inflation with the same exponent but
a different constant of proportionality.

\subsection{Perturbing the equilibrium position}
\label{sec:zdot}

We now relax the assumption that the second brane has to be at
constant $z$ and perturb the brane from its equilibrium position
found above. The governing equations are derived from the junction
conditions and prove not to be analytically soluble, though a
numeric solution can be obtained. Numeric solutions shown here
have initial conditions close to, but slightly perturbed from the
constant $z$ brane position.

\subsubsection{Junction Conditions}

The power-law inflation demonstrated in Section \ref{sec:prevres}
is induced by the action of the bulk scalar field on the brane,
with the brane being empty apart from the scalar field potential
(\ref{scalarfieldpotentialbr}). We now allow for matter with
energy density $\rho$ and pressure $p$ on the second brane.  Then
we find the junction conditions and the scalar field matching
conditions (given generally in \cite{Mukohyama&Coley}) at
the second brane become:

\begin{eqnarray}
\lefteqn{
\left[1-(H_0t\dot{z}_1)^2\right]^{-3/2}e^{-W}\Big[H_0t\ddot{z}_1 -
H_0^3t^2\dot{z}_1^3 } \nonumber \\ && \quad - W'H_0t\dot{z}_1^2 +
2H_0\dot{z}_1 +\frac{W'}{H_0t} \Big] \label{jcondn1}
\\
&=&-\frac{\kappa^2}{6}\{\lambda_1(\Phi)-2\rho-3p\} \, , \notag
\\
\lefteqn{
\left[1-(H_0t\dot{z}_1)^2\right]^{-1/2}e^{-W}\left[\frac{W'}{H_0t}
+ \frac{2H_0\dot{z}_1}{(3\Delta+8)}\right]}\notag
\\ &=& -\frac{\kappa^2}{6}\{\lambda_1(\Phi)+\rho\} \, , \label{jcondn2}
\\
\lefteqn{
\left[1-(H_0t\dot{z}_1)^2\right]^{-1/2}e^{-W}\Big[\frac{W'}{H_0t}}\notag
\\ && \quad+ \frac{2 H_0\dot{z}_1}{(3 \Delta+8)} \Big] \notag
\\ &=&
-\frac{\kappa^2}{6}\lambda_1(\Phi) \, . \label{jcondn3}
\end{eqnarray}

We can see that (\ref{jcondn2}) and (\ref{jcondn3}) are
inconsistent unless $\rho=0$.  The bulk metric ansatz and the form
of the scalar field matching condition we have chosen do not allow
for matter on the brane.  It would be possible to add matter by
taking a more general metric ansatz, or alternatively allowing a
coupling of the scalar field to matter to change the form of
equation \ref{jcondn3} (as discussed in \cite{LRM}).  Here,
however, we will investigate the case with no matter on the
branes.

When $\rho$ is taken to be zero, equations (\ref{jcondn2})
and (\ref{jcondn3}) are equivalent, and equation (\ref{jcondn1})
is equivalent to the derivative of these two. 

To evaluate the effect of any deviation from the
$z_1=const$ scenario previously considered we make the
perturbation $\overline{z}_1 \rightarrow \overline{z}_1+\delta z$,
where $\overline{z}_1$ is that value of $z_1$ that satisfies
equation (\ref{tuningreln}) and $\delta z$ is small.  Then the
linearised expansion of equation (\ref{jcondn2}) gives: \be
\frac{1}{H_0 t} \delta(e^WW') +
\frac{2H_0e^{-W}}{3\Delta+8}\delta\dot{z} =
\frac{2b}{6\kappa}\sigma_1\frac{\Xi'e^{-\sqrt{2}b\kappa\Xi}}{H_0t}\delta
z \, .\ee when we substitute the solution for $W$ and $\Xi$ given
(in the case $\Lambda<0$) by equations (\ref{exactmetric3}) --
(\ref{mathcalH}), this simplifies to \be
-\frac{3(\Delta+2)}{(3\Delta+8)}\frac{\delta z}{t} = \delta\dot{z}
\, . \ee  This has solution $\delta z \propto
t^{-3(\Delta+2)/(3\Delta+8)}$.  For the range of $\Delta$ we are
considering, the exponent of $t$ is positive, and hence the radion
is unstable.  We note that this solution agrees with the inverse
fourier transform of equation (49) of reference
\cite{Mukohyama&Coley} in the case where all bulk metric
perturbations are zero.

\subsubsection{A numeric solution for the brane trajectory}

To go beyond a linear analysis we solve equation
(\ref{jcondn2}) for $\dot{z}_1$:

\begin{eqnarray}
\dot{z}_1 &=&
\frac{72(3\Delta+8)}{72H_0^2\mathcal{H}^2+\sigma_1^2(3\Delta+8)^2}
\frac{\mathcal{H}}{t}
\Bigg[-\frac{\mathcal{H}'}{3(\Delta+2)}
\notag
\\*[-2mm] \label{dez1} \\* &&  - \frac{\sqrt{2}}{12} \sigma_1 \sqrt{1 +
\frac{\sigma_1^2(3\Delta+8)^2}{72H_0^2\mathcal{H}^2} -
\frac{(3\Delta+8)^2}{9(\Delta+2)^2}\frac{\mathcal{H}'^2}{H_0^2\mathcal{H}^2}}
\; \Bigg] \, . \notag
\end{eqnarray}

We introduce coordinates normalised with respect to reference
brane tension $\sigma_0$: \be \notag H \rightarrow
\frac{H}{\sigma_0} \, , \; \sigma_1 \rightarrow
\frac{\sigma_1}{\sigma_0} \, , \;  t \rightarrow \sigma_0 t \, ,
\; z \rightarrow \sigma_0 z \, . \ee In order to obtain a numeric
solution of the differential equation (\ref{dez1}) we choose
values for the parameters as: \bea \notag \Delta=-\frac{9}{4} \, ,
\; \delta=\frac{1}{4} \, , \; H=\frac{1}{12} \, ,  \\
H_0=\frac{5}{36} \, , \;  \sigma_1=-2 \, . \eea These have been
chosen to be consistent with the definitions and inequalities
presented in Section \ref{sec:prevres}.  Then the position of the
reference brane is given by
\begin{equation} z_0 = 12\sinh^{-1}{(2\sqrt{2})} \simeq 21.2 \, ,
\end{equation}
and the tuned second brane position giving constant $z_1$ is (from
equation (\ref{tuningreln}))
\begin{equation}\bar{z}_1 = 12\cosh^{-1}{(6)} \simeq 29.7 \, .
\label{tuningreln2}
\end{equation}
From equations (\ref{mathcalH}) and (\ref{dez1}) we get the
equation for $z_1(t)$:
\begin{equation} \dot{z}_1= \frac{36}{5} \frac{\left[\sinh{(z_1/12)}\cosh{(z_1/12)} -
6\sqrt{35} \, \right]}{\left[\{\cosh{(z_1/12)}\}^2+35\right]t} \,
. \label{dez1.2}
\end{equation}

\begin{figure}[hb!]
\begin{center}
\includegraphics[height=3in,width=3in]{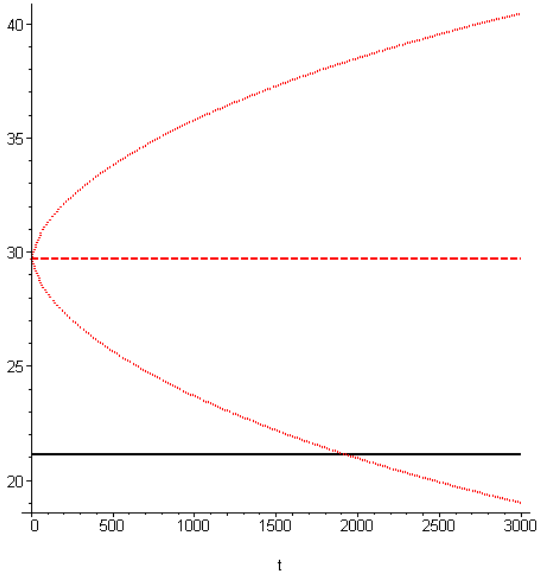}
\caption{Second brane position $z_1(t)$ plotted against bulk
coordinate time $t$.  Initial conditions are $z_1(0)=\bar{z}_1$
(dashed line) and $z_1(0)=\bar{z}_1\pm 0.1$ (dotted lines). The
reference brane position is also shown (solid line). The solution
for a brane perturbed towards the reference brane leads to a
collision at time $t_c \simeq 1920$, or brane proper time $\tau_c
\simeq 919$.} \label{plz1}
\end{center}
\end{figure}

Figure \ref{plz1} shows numeric solutions of this equation with
initial conditions chosen so that the three curves show the
trajectory of the second brane when it is initially at or slightly
to either side of, the tuned position calculated above in equation
(\ref{tuningreln2}).  The reference brane position is also shown,
and one can see that when the second brane is perturbed
\emph{towards} the reference brane, they will eventually collide.

\subsubsection{Scale factor on the moving brane}

As shown in \cite{SMS} the five-dimensional Einstein equations can
be projected onto a brane to obtain four-dimensional effective
equations. In the case of a bulk scalar field with no matter on
the branes, the effective four-dimensional equations on the brane
are~\cite{MaedaWands}:
 \be ^{(4)}G_{\mu \nu} =
-\frac{\delta}{2}\sigma_0^2e^{-2\sqrt{2}b\kappa\Phi} +
\frac{2}{3}\kappa^2T_{\mu \nu}(\Phi) - E_{\mu \nu} \, ,
\label{4dEFEs} \ee where $T_{\mu\nu}(\Phi) = D_{\mu}\Phi D_{\nu}
\Phi -\frac{5}{8}q_{\mu \nu}(D\Phi)^2$, the covariant derivative
on the brane is $D_\mu$ and the induced metric on the brane is
$q_{\mu \nu}$. $E_{\mu \nu}$ is the projection of the bulk Weyl
tensor onto the brane.

The Weyl part of these effective Einstein equations can be broken
up into a piece dependent on $\Phi$ and a piece independent of it
as follows \cite{Fphi}: \bea - E_{\mu \nu} &&= \sqrt{2}b\kappa
[D_{\mu}\Phi D_{\nu} \Phi - q_{\mu \nu}D^2\Phi]  \notag \\ && +
2b^2\kappa^2[D_{\mu}\Phi D_{\nu} \Phi - q_{\mu \nu}(D\Phi)^2]  \notag \\
&& + \frac{\kappa^2}{3}[D_{\mu}\Phi D_{\nu} \Phi -
\frac{1}{4}q_{\mu \nu}(D\Phi)^2] \notag \\ && +
\frac{6b^2}{\Delta}\delta\sigma_0^2e^{-2\sqrt{2}b\kappa\Phi}q_{\mu
\nu} + F_{\mu \nu} \, . \eea

$E_{\mu \nu}$ is necessarily traceless, giving \be
F_{\mu}^{\mu}=3\sqrt{2}b\kappa F_{\Phi} \ee  where
$3\sqrt{2}b\kappa F_{\Phi}$ is the trace of the $\Phi$ part of
$E_{\mu \nu}$, given by \be F_{\Phi} = D^2\Phi +
\sqrt{2}b\kappa(D\Phi)^2 -
4\sqrt{2}\frac{b\sigma_0^2\delta}{\kappa\Delta}e^{-2\sqrt{2}b\kappa\Phi}
\, . \ee

The bulk metric considered here is further constrained to have
$F_{\Phi}=0=F_{\mu\nu}$, where $F_{\Phi}=0$ gives the wave
equation for $\phi$.

The symmetries of the bulk metric combined with the wave equation
for the scalar field determine the induced metric and scalar field
on any brane in the bulk metric.  There are insufficient degrees
of freedom to allow for several independent solutions, with the
result that these quantities will have the same $\tau$-dependence
on a moving brane as on a fixed one.  This result is demonstrated
below using our numeric solution for the brane trajectory.

From the differential equation (\ref{dez1.2}) and the form of the
bulk metric we obtain the induced metric along the brane
trajectory as:
\begin{eqnarray} ds_{(4)}^2 &=& -d\tau^2 +256
[\sinh(z_1/12)]^{-16/3}\left(\frac{5 t}{36}\right)^{16/5}
d\vec{x}^2 \, , \notag
\\*[-4mm] \label{inducedmetric}
\end{eqnarray}
where proper time $\tau$ along the brane trajectory is related to
bulk coordinate time $t$ by
\begin{equation}\frac{d\tau}{dt} =
\frac{16}{[\sinh{(z_1/12)}]^{8/3}}
\frac{6\cosh{(z_1/12)}+\sqrt{35}\sinh{(z_1/12)}}{[\{\cosh{(z_1/12)}\}^2+35]}
\, . \label{detaut}\end{equation} Choosing the numeric solution
for $z_1$ with the second brane moving towards the reference
brane, equation (\ref{detaut}) can be solved numerically and
inverted to get $t(\tau)$.

From the induced metric (\ref{inducedmetric}) we have that the
scale factor $a_1$ on the second brane is
\begin{equation} a_1 = \frac{16}{[\sinh(z_1/12)]^{8/3}}\left(\frac{5t}{36}\right)^{8/5} \, ,\end{equation}
and, using (\ref{detaut}), we have \be H_1 =
\frac{1}{a_1}\frac{da_1}{d\tau} =
\frac{\sqrt{35}}{10t}[\sinh(z_1/12)]^{5/3} \, .\ee

\begin{figure}[ht!]
\begin{center}
\includegraphics[height=3in,width=3in,angle=-90]{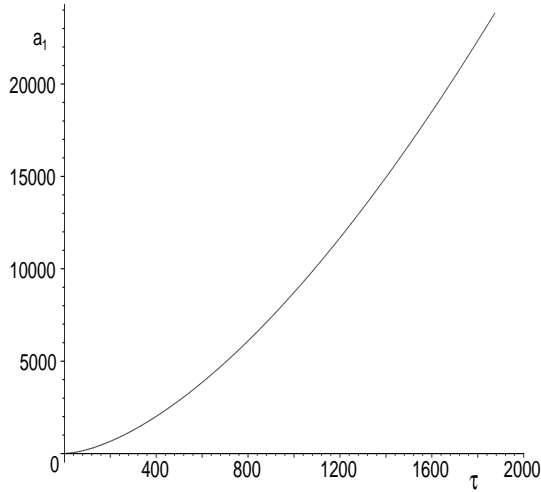}
\caption{Scale factor $a_1$ on the second brane plotted against
brane proper time $\tau$ for the trajectory perturbed towards the
reference brane.  We can see that there is no singularity or other
feature around the collision point $\tau_c \simeq 919$.}
\label{pl4a1}
\end{center}
\end{figure}

A plot of $a_1(\tau)$ is shown in Figure \ref{pl4a1}.  A
calculation of the quantity $\tau H_1(\tau)$ in the numeric
solution shows that it is constant at a value of
$2/(3\Delta+8)=8/5$, demonstrating that the scale-factor is
power-law with the same exponent as the non-moving brane.

\subsection{Constraints on the bulk metric}

In the above we have used a particular constrained form of the
bulk metric whose symmetry properties discounting any possibility
that the brane cosmology or the movement of the brane through the
bulk affect the form of the bulk metric.  In particular, the bulk
scalar field and the bulk Weyl tensor are constrained to symmetric
forms.  This symmetry may be described in terms of the two
quantities $F_\phi$ and $F_{\mu \nu}$ defined above.

Specifying the bulk metric ensures that this symmetry will not be
disturbed if the brane is perturbed and starts to move.  This
results in a first order evolution equation for the radion, where
other works have found second order equations~\cite{radion}. Given
an initial value for the radion, the brane velocity is fixed by
the evolution equation.  To set arbitrary initial position and
velocity, we must allow for some perturbation of the bulk metric
parameters.

Solving the full system of Einstein equations for a moving brane
would be far too complicated to be practical, but if we use a low
energy expansion we can examine the possibility for a more general
bulk in the low energy regime.

\section{Low Energy Expansion}\label{sec:lee}

The 4-dimensional effective theory valid at low energies for a
system with 2 branes is derived in \cite{LE1,Shiromizu:2002qr} and
applied to a bulk scalar field in \cite{LE2}. We will substitute
the potentials from equations (\ref{bulkpotential}) and
(\ref{scalarfieldpotentialbr}) into this low energy formalism to
determine the validity of allowing free brane motion while
constraining the bulk to the static metric given in \cite{koyama}.

In this low energy theory, there are essentially three moduli
fields $\hat{\alpha}$, $\hat{\phi}$ and $d$.  $\hat{\alpha}$ and
$\hat{\phi}$ correspond to the scale-factor and the scalar field
on the reference brane, and $d$ encodes an arbitrary variation of
the proper distance between the branes; the proper distance being
given by the formula

\be L= e^{\sqrt{2}b\hat{\phi}}d(t) \label{properdistance} \, . \ee

In this section we will use the variables and parameters of
\cite{LE2}, but to ensure we are describing the physical system
set up previously, the parameters must be given by
$\sigma=\sigma_0$, $V(\phi)=-2\delta\sigma^2e^{-2\sqrt{2}b\phi}$,
$U(\phi)=0$ and
$\tilde{U}(\phi)=\sigma'\sqrt{2}/\kappa^2e^{-\sqrt{2}b\phi}$ where
$\sigma'=\sigma_0-\sigma_1$; then the action given in \cite{LE2}
reduces to that of equation (\ref{action}).

The low energy expansion given in \cite{LE2} relies on the
assumption \be \sigma \gg \kappa^2\rho, \, \kappa^2 p, \,
\sigma^{-1}V, \, \kappa^2U, \, \kappa^2 \tilde{U} \, , \ee which
in our case implies \be \delta \ll 1 , \; \sigma' \ll \sigma \, .
\label{leineq}\ee In the $d=d_*=$ constant, $z_1=$ constant case
we can derive the coordinate transformation between the exact
metric given by equations
(\ref{exactmetric1})--(\ref{exactmetric4}) and the approximate one
of \cite{LE2} as: $ x\rightarrow x \, , \quad t\rightarrow t $ and
\bea \lefteqn{y \rightarrow
\frac{12}{\sqrt{2}(3\Delta+8)d_*\sigma}\Bigg[1} \notag
\\*[-2mm] \\*
&&-\left(\sqrt{-1-\frac{\Delta}{8\delta}}\sinh{(Hz)}\right)^{\frac{(3\Delta+8)}{3(\Delta+2)}}\Bigg]
\notag \eea (valid at low energies and small $\delta$ only). From
this transformation and the low energy conditions (\ref{leineq}),
we can derive the condition on the exact solution that $Hz<Hz_1\ll
1$. We suppose that this condition would also hold when $d$ is not
constant.

\subsection{Linear approximation}
\label{sec:lelin}

The effective equations on the brane are given by equations
(58)-(60) in \cite{LE2} . To evaluate the effect of some small
deviation from the $z_1=$ constant, $d=$ constant scenario we make
a linear perturbation $d=d_* + \delta_d$ where $\delta_d$ is small
compared to $d_*$.  We assume that $\sigma'$ is the tuned brane
tension to keep $d$ at constant $d_*$:
\be\sigma'=\frac{4\delta\sigma}{\Delta}\left[(1-d_*)^{(6\Delta+12)/(3\Delta+8)}-1\right]
\ee and that any deviation of $\hat{\phi}$ and $\hat{\alpha}$ from
their values in the exact solution is also of order $\delta_d$ or
smaller. Under these assumptions, the evolution equation for $d$
decouples from the governing equations for $\hat{\phi}$ and
$\hat{\alpha}$, and gives a second order ordinary differential
equation, \be \ddot{\delta}_d +
\left(\frac{3\Delta+14}{3\Delta+8}\right)\frac{\dot{\delta_d}}{t}
-
\left[\frac{9\Delta(\Delta+2)}{(3\Delta+8)^2}\right]\frac{\delta_d}{t^2}
= 0 \, . \label{deltad} \ee
 Substituting $F_{\phi}=0$ into the
evolution equation for $\phi$ obtained in the low energy expansion
and combining with the radion evolution equation (\ref{deltad})
gives a first order equation \be
\dot{\delta}_d=-\frac{3(\Delta+2)}{(3\Delta+8)}\frac{\delta_d}{t}
\label{Fphi0} \, . \ee

The second order evolution equation (\ref{deltad}) has two
independent solutions: \bea \delta_d &\propto&
t^{-3(\Delta+2)/(3\Delta+8)} \, , \label{growing} \\ \delta_d
&\propto& t^{3\Delta/(3\Delta+8)} \, . \label{decaying} \eea The
growing mode (\ref{growing}) satisfies the $F_{\phi}=0$ condition
(\ref{Fphi0}), while the decaying mode (\ref{decaying}) does not.
If one chooses initial conditions for the brane motion satisfying
$F_{\phi}=0$, they specify solely the growing mode, so a solution
that starts with $F_{\phi}=0$ will remain that way for all time.
The one-parameter family of solutions Section \ref{sec:zdot} must
lie along this growing mode trajectory. Any solution that also
contains some decaying mode solution will not have $F_{\phi}=0$,
but it will tend towards a trajectory that does, implying that
$F_{\phi}=0$ is in some sense an attractor.

\begin{figure}[hbt]
\begin{center}
\includegraphics[height=3in,width=3in]{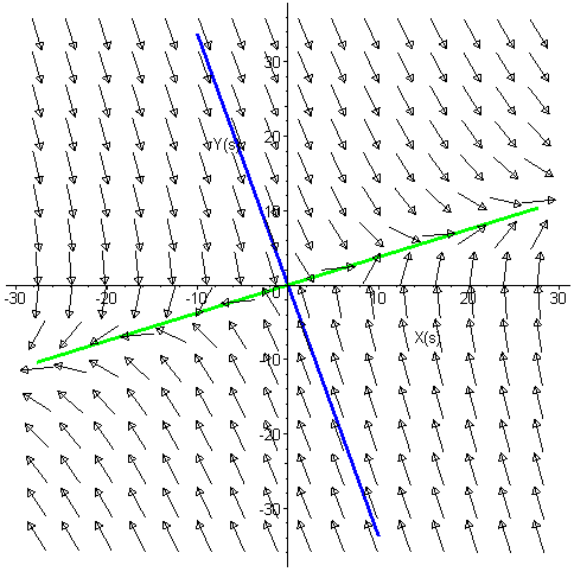}
\caption{Phase plane for the linearised evolution of $d$ when
$\Delta=-\frac{9}{4}$. The line with positive gradient corresponds
the growing mode for d shown in equation (\ref{growing}), and the
line with negative gradient corresponds to the decaying mode
(equation (\ref{decaying})) } \label{phaseplane}
\end{center}
\end{figure}

To demonstrate more clearly the nature of the solutions to the
linearised evolution equation for $d$, we present a phase-plane
analysis.  To transform equation (\ref{deltad}) to an autonomous
form, we make a change of variable to $s=\frac{2}{(3\Delta+8)}\ln
t$.  Using this new time coordinate, the evolution equation
becomes \be \delta_d '' + 3\delta_d '
-\frac{9}{4}\Delta(\Delta+2)\delta_d = 0 \, , \ee where a prime
denotes the derivative with respect to s.  If we put $X=\delta_d$
and $Y=\delta_d '$, then the system can be written \be X' = Y \, ,
\; Y' = \frac{9}{4}\Delta(\Delta+2)X - 3Y \, . \label{linsys} \ee
The system has only one equilibrium point, at $X=Y=0$.  The
eigenvalues of the system around this point are
$\frac{3\Delta}{2}$ and $\frac{-3(\Delta+2)}{2}$, with the first
negative and the second positive.  Thus $X=Y=0$ is a saddle point.
The equilibrium lines of this saddle point are found to be
$Y=\frac{3\Delta}{2} X$ and $Y=\frac{-3(\Delta+2)}{2} X$.  The
equations (\ref{linsys}) show that the phase trajectories go from
left to right, i.e. towards the equilibrium line with positive
gradient. Converting the equilibrium lines back to our original
coordinates confirms that they correspond to the solutions
(\ref{growing}) and (\ref{decaying}) respectively, and that the
growing mode (\ref{growing}) is an attractor line.  The phase
diagram for $\Delta=-\frac{9}{4}$ is shown in Figure
\ref{phaseplane}.

Earlier we made the assumption that $\hat{\alpha}$ and
$\hat{\phi}$ differed only by a small amount from the background
solution. We now quantify this as \be
\hat{\alpha}=\frac{2\ln(H_0t)}{(3\Delta+8)}+\delta_{\alpha}(t) \,
, \; \hat{\phi}=\frac{\ln(H_0t)}{\sqrt{2}b}+\delta_{\phi}(t) \ee
where $\delta_{\alpha}$ and $\delta_{\phi}$ are small relative to
the background. Now the linear reduction of the remaining
evolution equations from \cite{LE2} is obtained as: \bea
\lefteqn{\ddot{\delta}_{\phi} +
\frac{3}{\sqrt{2}b}\frac{\dot{\delta}_{\alpha}}{t} +
\frac{(6\Delta+22)}{(3\Delta+8)}\frac{\dot{\delta}_{\phi}}{t} +
\frac{12}{(3\Delta+8)}\frac{\delta_{\phi}}{t^2}} \notag \\*[-2mm]
\label{deltaphi}
\\*[-2mm] &=& \notag
\frac{-18\sqrt{2}b(\Delta+4)}{(3\Delta+8)^2}\xi(d_{*})\left[\frac{\dot{\delta}_d}{t}
+ \frac{3(\Delta+2)}{(3\Delta+8)}\frac{\delta_d}{t^2} \right] \, ,
\\[4mm]
\lefteqn{\ddot{\delta}_{\alpha} +
\frac{8}{(3\Delta+8)}\frac{\dot{\delta}_{\alpha}}{t} +
\frac{1}{3\sqrt{2}bt}\frac{\dot{\delta}_{\phi}}{t} -
\frac{6\Delta\sqrt{2}b}{(3\Delta+8)^2}\frac{\delta_{\phi}}{t^2}}
\notag \\*[-2mm] \label{deltaalpha}  \\*[-2mm] \notag &=& 0 \, ,
\\[4mm]
\lefteqn{\frac{\dot{\delta}_{\alpha}}{t} +
\frac{2\sqrt{2}b}{(3\Delta+8)}\frac{\delta_{\phi}}{t^2}} \notag
\\*[-2mm] \label{deltaconstraint}  \\*[-2mm] \notag &=&
\frac{-2}{(3\Delta+8)}\xi(d_*)\left[\frac{\dot{\delta}_d}{t} +
\frac{3(\Delta+2)}{(3\Delta+8)}\frac{\delta_d}{t^2}\right] \, ,
\eea where \be
\xi(d_*)=\frac{(1-d_*)^{4/(3\Delta+8)}}{\{1-(1-d_*)^{(3\Delta+12)/(3\Delta+8}\}}
\notag \ee

The system (\ref{deltad}), (\ref{deltaphi}), (\ref{deltaalpha}),
(\ref{deltaconstraint}) has a general solution of the form \bea
\delta_{d} &=& d_1 t^{-3(\Delta+2)/(3\Delta+8)} + d_2
t^{3\Delta/(3\Delta+8)} \, ,
\\[2mm]
\delta_{\phi} &=& \phi_1 t^{-1} + \phi_2 t^{-6/(3\Delta+8)} \notag
\\*[-3mm] \\*[-2mm] \notag && \; -
d_2\xi(d_*)\frac{3(\Delta+1)}{\sqrt{2} b(3\Delta+4)}t^{3\Delta/(3\Delta+8)}
\, ,
\\[2mm]
\delta_{\alpha} &=&
\frac{2\sqrt{2}b}{(3\Delta+8)}\left[\phi_1t^{-1}
+\frac{(3\Delta+8)}{6}\phi_2t^{-6/(3\Delta+8)}\right] \notag
\\* \\*[-4mm] \notag \; && -
d_2\xi(d_*)\frac{6(\Delta+1)}{(3\Delta+8)(3\Delta+4)}t^{3\Delta/(3\Delta+8)}
+ c_{\alpha} \, , \eea
 where $d_1$, $d_2$, $\phi_1$, $\phi_2$ and
$c_{\alpha}$ are all arbitrary constants of integration.  However,
we can neglect $c_{\alpha}$ as it is a gauge mode corresponding to
a redefinition of the ordinary spatial coordinates. We can then
see from the form of this solution that any perturbation to the
background metric functions $\hat{\alpha}$, $\hat{\phi}$ will
decay (in the linear regime at least).

\subsection{Full non-linear low energy evolution}

\begin{figure}[hbt]
\begin{center}
\includegraphics[height=3in,width=3in]{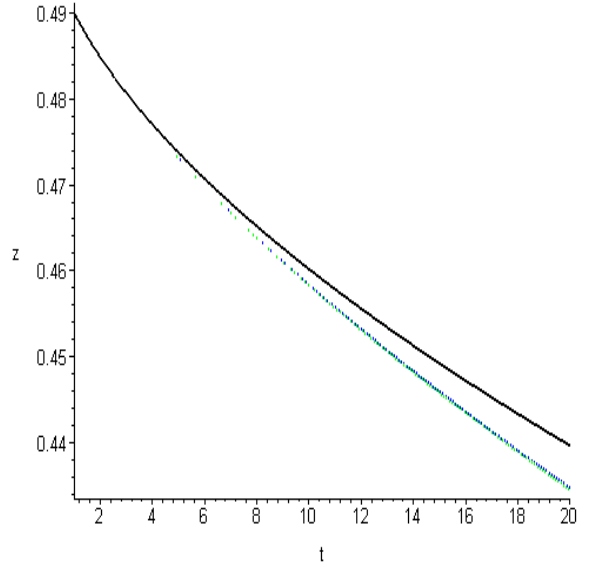}
\caption{Numerical simulations of the non-linear radion evolution
(dotted lines) are compared to the solution for the radion in the
linear regime (solid line)} \label{radiongraph}
\end{center}
\end{figure}

Having completed the linear analysis, we solve the full non-linear
system of low energy equations numerically, to see if the results
above are substantially changed in the non-linear regime. The
parameters chosen for this numerical analysis are \be
\Delta=-\frac{9}{4}\, , \; \delta=\frac{1}{100}\, , \;
H=\frac{1}{60}\, , \; H_0=\frac{1}{36} \ee (the value of $\delta$
chosen for the previous numerical simulation was too large to be
valid in the low energy approximations, so more appropriate values
have been chosen).  The background value of $d$ is taken to be
$d_*=0.5$ and the corresponding tuned value of the potential on
the second brane is $\sigma' \simeq -0.023$.  Figure
\ref{radiongraph} shows the numerical analysis (dotted curves)
compared to the solution for the radion evolution from the linear
regime, shown as the solid curve. We can see that at early time
our linear approximation is reasonably accurate, but that as the
collision is approached the two curves are widely divergent.

The dotted curves show the results for the numerical simulation
with two different sets of initial conditions - one set of initial
conditions correspond to the growing mode identified in the linear
analysis, and hence to the brane trajectories identified in the
exact solution, whereas the other set correspond to a mixture of
the growing and decaying modes.  The numerical solution for
$\hat{\alpha}$ and $\hat{\phi}$ under these initial conditions
shows that while they have identically the form of the background
solution for the growing-mode conditions, the mixed-mode initial
conditions set an initial perturbation away from the background
solution which quickly decays to $0$ (or to the constant gauge
mode in the case of $\hat{\alpha}$).  This confirms the stability
of the background solution to perturbations of the brane position
that was demonstrated in the linear regime.

\subsection{Accuracy of the low energy approximation and simulations}

\begin{figure}[htb]
\begin{center}
\includegraphics[height=3in,width=3in]{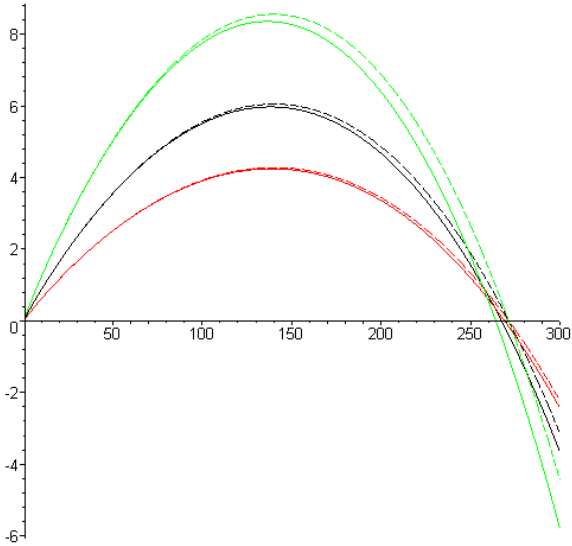}
\caption{The proper distance between the two branes as given by
the exact solution (solid lines) and low-energy non-linear
equations (dashed lines), plotted for $\delta=1/200$ (lowest),
$\delta=1/100$ (middle) and $\delta=1/50$ (highest).}
\label{deltasens}
\end{center}
\end{figure}

\begin{figure}[htb]
\begin{center}
\includegraphics[height=3in,width=3in]{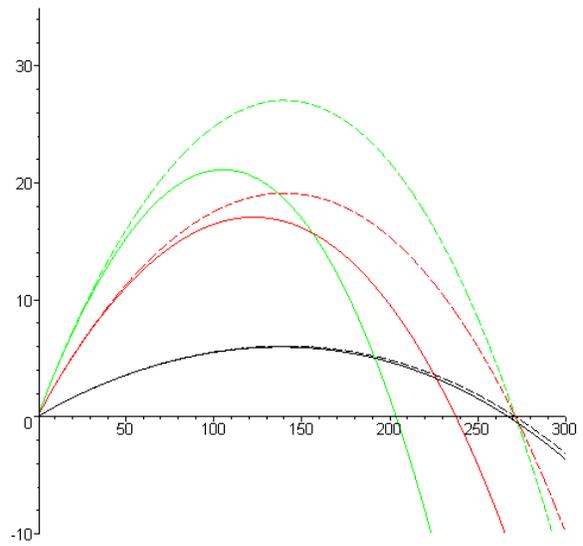}
\caption{The proper distance between the two branes as given by
the exact solution (solid lines) and low-energy non-linear
equations (dashed lines), plotted for $\delta=1/5$ (highest),
$\delta=1/10$ (middle) and $\delta=1/100$ (lowest)}
\label{deltasens2}
\end{center}
\end{figure}

Figures \ref{deltasens} and \ref{deltasens2} show the proper
distance between the two branes (given in equation
(\ref{properdistance})) as obtained from the numerical integration
of the low energy effective equations and from the numerical
solution for the brane trajectory in the exact bulk metric
solution, plotted for several different values of the parameter
$\delta$. There is some difference between the exact and low
energy curves, but we can see this difference decreasing as
$\delta$ decreases, indicating that the integration of the low
energy equations is accurate in the limit $\delta \rightarrow 0$.
Figure \ref{deltasens} demonstrates the sensitivity of the system
to even small changes in $\delta$, whereas Figure \ref{deltasens2}
demonstrates the wide deviation of the approximate solution from
the exact solution when $\delta$ becomes larger.

\begin{figure}[hbt]
\begin{center}
\includegraphics[height=3in,width=3in]{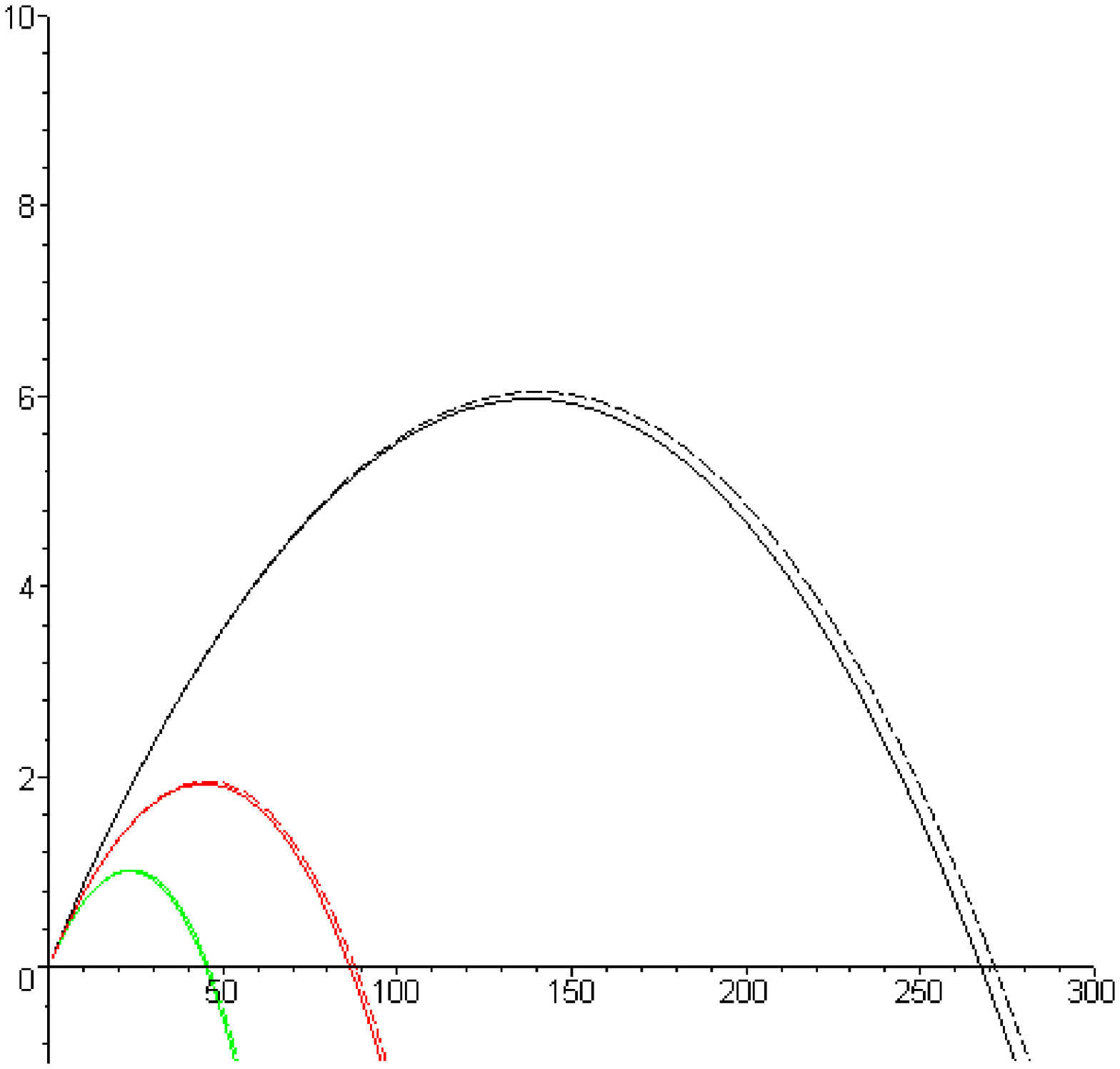}
\caption{The proper distance between the two branes as given by
the exact solution (solid lines) and low-energy non-linear
equations (dashed lines), plotted for a variety of initial
perturbations of $d$ ($\delta_d=0.01$, $0.02$, $0.03$).  Time to
the collision decreases with increasing initial perturbation.}
\label{inisens}
\end{center}
\end{figure}

Figure \ref{inisens} demonstrates the sensitivity of both
solutions to a change in initial conditions.  An increase in the
initial perturbation in the distance between the branes
dramatically decreases the time to the brane collision.

\section{Conclusions}

In this paper we have demonstrated that for the bulk metric with
scalar field given in \cite{koyama} there is an instability in the
position of a second brane.  Solving for the trajectory of this
second brane shows that the instability can lead to a collision
between the two branes.  The scale factor and scalar field are
shown to retain the same dependence on brane proper time
regardless of the motion of the brane, and hence remain regular at
the collision.  It was found in \cite{websterdavis,Martin:2003yh}
that a general class of potentials for 2-brane systems with bulk
scalar field will result in a singularity at any collision;
however it was found in \cite{Martin:2003yh}, as here, that the
exponential potential is exceptional and can tame the
singularities at the collision.

It is also worth noting that (for a countably infinite subset of
the possible values of the parameters $b$ and $\Delta$) the scalar
field potential used here can be reproduced from a
compactification of an empty higher dimensional space
\cite{cveticlupope,kobayashitanakaxtradim}, with the
dimensionality determining the values of $\Delta$ and $b$.  In
this picture the collision is singular in that the bulk
spacetime loses a dimension when the space between the brane
disappears, however the induced metric of the brane and
compactified space remains regular, and hence the effective scalar
field derived from the compactified volume will remain regular
also.

In a complementary approach we considered the system using a low
energy expansion developed in \cite{LE2}. This expansion
reproduces the solution found in the previous section to good
accuracy in the correct limit (see Figure \ref{deltasens}).  The
low energy equations also identify a solution excluded by the
methods used previously, in which the initial perturbation of the
second brane necessitates a perturbation of the bulk metric away
from its background.  This solution quickly decays toward the
original solution with the unperturbed bulk, confirming the
conclusion of \cite{Mukohyama&Coley} that there can be no unstable
homogeneous perturbations of the bulk metric.

It would be interesting to see if the capability of solving the
cosmological perturbations demonstrated in Koyama and Takahashi's
one-brane solution \cite{koyama2} is reproduced in this two-brane
model.  The perturbations would diverge as the branes approach collision,
but it is possible to match divergent perturbations across a collision in
some scenarios (see e.g. \cite{bornagain}).  If this is possible it would
give an interesting parallel to the
``Born-again braneworld" \cite{bornagain}, the cyclic model
\cite{cyclic} and other pre-big-bang scenarios.  A microscopic
particle-theoretic understanding of interactions at the collision
would be helpful in determining the dynamics at the collision
(whether the branes are likely pass through each other as in
\cite{bornagain,cyclic}, coalesce as in
\cite{KannoSodaWands,JamesGray} or disintegrate as in
\cite{eppur}), as well as the behaviour of perturbations near the
collision.  However there are techniques available to evolve
perturbations through brane collisions at a purely classical level
\cite{bornagain,tolleyturoksteinh}.

\section{Acknowledgements}

EL is supported by EPSRC; KK and RM are supported by PPARC.

\end{document}